\documentclass[aps,prd,floatfix,superscriptaddress,preprintnumbers]{revtex4}
\usepackage{amssymb}
\usepackage{amsmath}
\usepackage{amsfonts}
\usepackage{orcidlink}
\usepackage{epsfig}             
\usepackage{graphicx}
\usepackage{stackrel}
\usepackage{tabularx}
\usepackage{color}

\newcommand{\eq}{\begin{eqnarray}}
\newcommand{\en}{\end{eqnarray}}

\begin{document}

\title{Quantum states in an AdS-invariant theory} 

\author{Felix~M.~Lev}
\email{felixlev314@gmail.com}
\affiliation{Institut f\"ur Theoretische Physik, Universit\"at T\"ubingen, \\
Kepler Center for Astro and Particle Physics, \\ 
Auf der Morgenstelle 14, D-72076 T\"ubingen, Germany} 

\author{Valery~E.~Lyubovitskij \orcidlink{0000-0001-7467-572X}}
\email{valeri.lyubovitskij@uni-tuebingen.de}
\affiliation{Institut f\"ur Theoretische Physik, Universit\"at T\"ubingen, \\
Kepler Center for Astro and Particle Physics, \\ 
Auf der Morgenstelle 14, D-72076 T\"ubingen, Germany} 
\affiliation{Millennium Institute for Subatomic Physics at
the High-Energy Frontier (SAPHIR) of ANID, \\
Fern\'andez Concha 700, Santiago, Chile}
\affiliation{Department of Physics, Tomsk State University, 634050 Tomsk, Russia}

\begin{abstract}

We note that the description of elementary particles in anti-de Sitter (AdS) 
quantum theory differs radically from standard approaches. Since the Cartan
subalgebra of the AdS algebra has rank 2, it is necessary to consider only
those elements that explicitly depend on the eigenvalues of operators in
the irreducible representations (IRs) of the algebra. In the literature,
AdS symmetry is widely invoked in various quantum applications, such as AdS/QCD.
However, such frameworks typically do not take into account the unconventional
properties of the AdS algebra IRs investigated in the present paper.

\end{abstract}

\maketitle

\section{Introduction}
\label{Intro}

In quantum theory, the state of a system is described by an element of
a Hilbert space, and the physical quantities characterizing the system
are associated with the eigenvalues of certain self-adjoint operators
in that space. It is well known that states corresponding to distinct
eigenvalues of a self-adjoint operator are mutually orthogonal.
This raises the following question. Let ${\cal A}$ be a set of
self-adjoint operators $\{A_1,A_2,...\}$, and let $\Psi_1$ and $\Psi_2$
be two {\it distinct} physical states such that
$A_i\Psi_j=\lambda_{ij}\Psi_j$ ($j=1,2$).
Suppose that $\Psi_1$ and $\Psi_2$ share the same set of eigenvalues,
i.e., $\lambda_{i1}=\lambda_{i2}$ for $\forall i$.
Under these conditions, the following assertion seems natural:

{\bf Conjecture.} {\it If the states $\Psi_1$ and $\Psi_2$ are distinct,
  then there exists a self-adjoint operator $B$ such that
  $\Psi_1$ and $\Psi_2$ are eigenvectors of this operator with distinct
  eigenvalues, i.e., if $B\Psi_j=\lambda_j\Psi_j$,
  then $\lambda_1 \neq \lambda_2$.} 

The history of quantum mechanics appears to validate this {\bf Conjecture}.
For example, if we consider two states of an elementary particle with different
three-momenta ${\bf p}_1 \neq {\bf p}_2$ differing by at least one spatial component,
the eigenvalues of the momentum operator $P_i$ for these states are distinct.
Similarly, for bound states of a particle with different energies,
the eigenvalues of the energy operator will differ. As a final example,
if we consider two distinct states of the same particle with different
angular momentum projections, the eigenvalues of the corresponding
projection operator will be different. These examples suggest that
the {\bf Conjecture} is intuitively obvious and requires
no formal justification.

This consensus reflects a deeply ingrained intuition among quantum physicists,
who often take for granted that distinct physical states must always be
distinguishable by at least one self-adjoint operator. However, to the best
of our knowledge, no rigorous theorem proving this {\bf Conjecture} has ever
been established in the literature. From a strictly mathematical standpoint,
if it is asserted that a distinguishing operator with specific spectral
properties exists, a constructive algorithm for finding it must be provided.
The lack of such a constructive proof indicates that the validity of the
{\bf Conjecture} is assumed implicitly rather than proven formally.

The history of physics demonstrates that when certain phenomena cannot
be accommodated within conventional frameworks, they must be described
on the basis of entirely new concepts. If we proceed from the premise
that quantum theory is rooted in Quantum Field Theory (QFT),
the symmetry of the theory is conventionally dictated by the underlying
spacetime background. For instance, non-relativistic theory relies
on Galilean spacetime symmetry; relativistic theory is governed
by Poincaré symmetry; and de Sitter (dS) or anti-de Sitter (AdS)
theories inherit the symmetries of dS or AdS spaces, respectively.
Although QFT has achieved monumental success, it suffers from
well-known foundational issues. In non-renormalizable theories
(such as local quantum gravity), divergences cannot be systematically
eliminated, hindering direct comparison with experiment.
Furthermore, the very existence of point-like elementary particles is
at odds with the presence of infinitesimals in standard mathematics.

As discussed in Ref.~\cite{book} and related publications,
QFT exhibits a notable conceptual inconsistency.
According to the principles of quantum mechanics,
every physical quantity must be represented by a self-adjoint operator.
However, no self-adjoint operators correspond to the coordinates of
the spacetime background in QFT. One might argue that since
the spacetime background serves merely an auxiliary role in
constructing physical operators, its coordinates need not be
directly measurable. Nevertheless, it remains unclear why
non-physical quantities must be introduced to construct
the theory if all physical operators can be formulated
without invoking them.

Alternatively, as elaborated in Ref.~\cite{book},
a consistent quantum theory can be constructed by defining
the symmetry of the system directly through the choice of
its operator algebra, rather than via a spacetime background.
Specifically, the Galilei algebra is chosen for non-relativistic
systems, the Poincaré algebra for relativistic ones, and
the dS/AdS algebras for dS/AdS-invariant systems.
By definition, an elementary particle is a system described
by a unitary irreducible representation (UIR) of the chosen
symmetry algebra in a Hilbert space. The term ``unitary''
is retained here for consistency with standard group-theoretic
terminology, though we focus purely on the algebra level.

This algebraic approach provides a rigorous criterion for
determining the generality of a given symmetry. For instance,
the transition from relativistic to non-relativistic theory
is understood as a mathematical contraction of the Poincaré
algebra to the Galilei algebra via $c\to\infty$~\cite{Dyson,book}.
Consequently, any non-relativistic result can be reproduced
in relativistic theory by choosing a sufficiently large value
of $c$, whereas non-relativistic theory cannot capture phenomena
where the finiteness of $c$ is essential.

In his seminal 1972 paper, Dyson~\cite{Dyson} demonstrated that
relativistic theory is a special case of dS and AdS theories,
as the Poincaré algebra is obtained from the dS or AdS algebras
via the contraction $R\to\infty$ (see also Ref.~\cite{book}),
where $R$ acts as a fundamental dimensionful parameter defining
the scale of the symmetry algebra. At the quantum level, this
structural contraction constant $R$ possesses the dimension of
inverse energy (or length) and has nothing to do with the geometric
curvature radius of classical dS or AdS spacetime backgrounds.
Instead, $R$ governs the non-commutativity of the translation-like
generators, establishing a soft constraint on the applicability of
flat Poincaré invariance to internal hadronic structures.
It follows that:
(a) any result of relativistic theory can be reproduced
in dS/AdS theories for a sufficiently large parameter $R$;
(b) relativistic theory fails to capture those effects
where the finiteness of $R$ is crucial. Since dS and AdS algebras
are semisimple, they represent the most general symmetry structures,
as they cannot be obtained via contraction from any other algebra.

Despite these fundamental insights established over 50 years ago,
systematic attempts to generalize quantum electrodynamics,
quantum chromodynamics, and the electroweak theory to quantum
dS or AdS symmetries have remained confined to the standard
geometric QFT framework and have not gained wide traction.
A key reason is that within QFT, $R$ is interpreted as the
cosmological radius of the Universe. Since this scale vastly
exceeds the characteristic sizes of elementary particles,
it is widely assumed that dS/AdS modifications are relevant
only in cosmology. However, as shown in Ref.~\cite{book} and
further demonstrated in this work, the algebraic effects of
dS and AdS symmetries are highly significant for
particle physics at the quantum level.

In this paper, we explicitly show that the description of
elementary particles in quantum AdS theory differs radically
from the standard relativistic description,
leading to a new, predictive framework for hadronic physics.
In particular, we show that the widely accepted {\bf Conjecture}
fails when applied to elementary particles within
an AdS invariant framework. Given that AdS symmetry
is more general than Poincaré symmetry~\cite{Dyson},
this necessitates a revision of our understanding of
the fundamental properties of quantum states.

The paper is organized as follows.
In Sec.~\ref{VS2}, we review the properties of the UIRs
of the $sp(2)$ algebra in a form convenient for constructing
AdS representations.
In Sec.~\ref{VS3}, we outline the algebraic properties of
the UIRs of the AdS algebra.
Sec.~\ref{SSEvans} reviews the standard mathematical techniques
for constructing these UIRs,
while Sec.~\ref{VS4} details why these approaches are not fully
compatible with the physical requirements of quantum theory.
In Sec.~\ref{Thermodynamics}, we investigate the thermodynamic consequences
of the proposed state-space restriction for a quantum gas with AdS symmetry.
Finally, our conclusions and future perspectives for non-perturbative hadronic
physics are outlined in Sec.~\ref{Conclusion}.

\section{UIRs of the sp(2) algebra}
\label{VS2}

A key role in constructing positive-energy UIRs of the AdS algebra
is played by the positive-energy UIRs of its $sp(2)$ subalgebra.
These are defined by a set of operators $(a',a'',h)$ satisfying
the commutation relations:
\eq\label{V2}
[h,a'] = - 2 a' \,, \quad [h,a''] = 2 a'' \,, \quad [a',a''] = h \,.
\en

The second-order Casimir operator $K$ of the $\mathfrak{sp}(2)$
subalgebra is defined as:
\eq\label{V3}
K = h^2 - 2 h - 4 a' a'' = h^2 - 6 h - 4 a'' a'
\,.
\en 
We consider UIRs defined on a lowest-weight vector $e_0$
with positive energies such that:
\eq\label{V4}
a' e_0  = 0 \,, \quad h e_0 = q_0 e_0 \,,
\en
where $q_0 > 0$ is an integer. Defining the basis vectors as
$e_n=(a'')^ne_0$ for $n=1,2,\ldots$, 
it follows from Eqs.~(\ref{V3}) and~(\ref{V4}) that:
\eq\label{V6}
h e_n = q_n e_n\,, \quad
K e_n = q_0 (q_0-2) e_n \,, \quad
a' a'' e_n = (n+1) (q_0+n) e_n
\,,
\en
where $q_n = q_0 + 2 n$. By construction, $q_n > 0$ and the corresponding UIR
is infinite-dimensional. Analogously, negative-energy UIRs can be constructed
starting from a highest-weight vector $e_0'$ satisfying:
\eq\label{V7}
a'' e_0' = 0 \,, \quad h e_0' = - q_0 e_0' 
\en
with the higher states defined as $e_n' = (a')^n e_0'$ ($n=1,2,...$).
In standard quantum theory, IRs are defined on Hilbert spaces equipped
with a positive-definite scalar product. This product is normalized such
that $(e_0,e_0)=1$, the operator $h$ is self-adjoint, and the operators
$a'$ and $a''$ are mutually adjoint, i.e., $(a')^*=a''$.
Then, from Eq.~(\ref{V6}), we obtain:
\eq\label{scalar}
(e_n,e_n) = n! \, (q_0)_n \,,
\en
where $(q_0)_n = q_0(q_0+1)\cdots (q_0+n-1)$ denotes the Pochhammer symbol,
which can be equivalently expressed in terms of the standard Euler
Gamma functions as:
\eq\label{pochhammer_gamma}
(q_0)_n = \frac{\Gamma(q_0 + n)}{\Gamma(q_0)} \,.
\en
Conventionally, positive-energy UIRs correspond to particles and negative-energy
UIRs to antiparticles, with the energies of the latter becoming positive
upon second quantization. Choosing $q_0$ as an integer is particularly
advantageous when generalizing the standard framework to a theory based
on finite rings, where both positive- and negative-energy UIRs coalesce
into a single representation~\cite{book}.
In this work, however, we restrict our focus to standard UIRs.
For comparison, the representations of the standard $su(2)$ algebra
are defined by the operators $(L_+,L_-,L_3)$,
where $L_\pm = (L_1 \pm i L_2)/2$, satisfying:
\eq\label{su2CR}
         [L_3,L_+]=2L_+  \,, \quad
         [L_3,L_-]=-2L_- \,, \quad
         [L_+,L_-]=L_3   \,.
\en
These are related to the standard Pauli matrices via
$L_\pm = (1/2) (\sigma_1 \pm i\sigma_2)$ and $L_3 = \sigma_3$.
The basis $(L_+,L_-,L_3)$ is used here as it is highly convenient
for spectral calculations.

\section{Unitary representations of the AdS algebra}
\label{VS3}

Unitary representations of the AdS algebra relevant
for particle physics have been extensively studied,
for instance, by Evans~\cite{Evans}.
The basis of these representations is formed by ten
self-adjoint operators $M^{ab}$ satisfying
$M^{ba}=-M^{ab}$ ($a,b=0,1,2,3,4$).
The commutation relations are written in the canonical forms:
\eq\label{newCR}
[M^{ab},M^{cd}]=-i (\eta^{ac} M^{bd}
                 + \eta^{bd} M^{ac}
                 - \eta^{ad} M^{bc}
                 - \eta^{bc} M^{ad}) \,,
\en
where the metric $\eta^{ab}$ has diagonal components
$\eta^{00}=\eta^{44}=-\eta^{11}=-\eta^{22}=-\eta^{33}=1$.
Following the contraction procedure $R\to\infty$~\cite{Dyson,book},
the generator $M^{04}$ acts as the AdS analog of the energy operator.
Equations~(\ref{newCR}) are expressed in the standard dimensionful units
where $\hbar = c = 1$. As emphasized in Ref.~\cite{book},
a fundamental quantum theory should ideally exclude $\hbar$ and $c$
from its core equations, and indeed Eq.~(\ref{newCR}) does not
fundamentally contain those quantities. The chosen canonical normalization
ensures that the minimum non-zero value of the angular momentum is $1/2$,
in strict alignment with standard quantum-mechanical conventions for
spin-1/2 fermions and integer-spin bosons.

To describe the UIRs explicitly, we adopt an alternative set of ten operators:
$(a_j',a_j'',h_j)$ with $j=1,2$, $(b',b'')$, and $(L_+,L_-)$, where $L_3=h_1-h_2$.
Here, the index $j=1,2$ explicitly enumerates two independent constituent
$sp(2)$ subalgebras. Let $(a_j',a_j'',h_j)$ be these two independent sets
of operators satisfying the $sp(2)$ commutation relations:
\eq\label{V9}
[h_j,a_j']=-2a_j'  \,, \quad
[h_j,a_j'']=2a_j'' \,, \quad
[a_j',a_j'']=h_j   \,.
\en
The sets are independent because for different values of the subalgebra
index $j$ they mutually commute with each other. For each $sp(2)$ subalgebra,
the corresponding second-order local Casimir operator $K_j$ ($j=1,2$)
is defined in accordance with Eq.~(\ref{V3}) as:
\eq\label{K_j_def}
K_j = h_j^2 - 2h_j - 4a_j' a_j'' = h_j^2 - 6h_j - 4a_j'' a_j' \,.
\en
The remaining commutation relations read:
\eq\label{V11a}
& &[a_1',b']=[a_2',b']=[a_1'',b'']=[a_2'',b'']=0  \,,
\nonumber\\
& &[a_1',b'']=[b',a_2'']=L_-     \,, \quad
   [a_2',b'']=[b',a_1'']=L_+     \,, \nonumber\\
& &[b',b'']=\frac{1}{4}(h_1+h_2) \,,
\en
\eq\label{V11b}
& &[a_1',L_-]=[a_1'',L_+]=[a_2',L_+]=[a_2'',L_-]=0  \,,
\nonumber\\
& &[a_1',L_+]=[a_2',L_-]=b'     \,, \quad
   [a_2'',L_+]=[a_1'',L_-]=-b'' \,,
    \nonumber\\
& &[b',L_-]=\frac{1}{2}a_1'               \,, \quad
   [b',L_+]=\frac{1}{2}a_2'               \,,
   \nonumber\\
& &[b'',L_-]=-\frac{1}{2}a_2''            \,, \quad
   [b'',L_+]=-\frac{1}{2}a_1''            \,,
\en
and
\eq\label{V11c}
& &[h_j,b']=-b' \,, \quad [h_j,b'']=b'' \,,
\nonumber\\
& &[h_1,L_{\pm}]=\pm L_{\pm} \,, \quad [h_2,L_{\pm}]=\mp L_{\pm}
\,.
\en
These relations arise naturally within the Weyl basis of the AdS algebra.
The connection to the standard generators $M_{ab}$ is given
by:
\begin{align}
\label{V12}
& M_{10}=\frac{i}{2}(a_1''-a_1'-a_2''+a_2') \,, \quad
  M_{14}=\frac{1}{2}(a_2''+a_2'-a_1''-a_1') \,,
\nonumber\\
& M_{20}=\frac{1}{2}(a_1''+a_2''+a_1'+a_2') \,, \quad
  M_{24}=\frac{i}{2}(a_1''+a_2''-a_1'-a_2') \,,
 \nonumber\\
& M_{12}=L_3          \,, \quad
  M_{23}=L_++L_-      \,, \quad
  M_{31}=-i(L_+-L_-)  \,,
 \nonumber\\
& M_{04}=\frac{1}{2}(h_1+h_2) \,, \quad
  M_{34}=b'+b''               \,, \quad
  M_{30}=-i(b''-b')           \,.
\end{align}
The global second-order Casimir invariant $I_2$ for this AdS representation
of the entire anti-de Sitter $\mathfrak{so}(3,2)$ algebra is defined
in the standard form as:
\eq\label{I2AdS}
I_2 = \frac{1}{2} \sum\limits_{ab} M_{ab} M^{ab} \,,
\en
which, via Eqs.~(\ref{su2CR}) and~(\ref{V9})-(\ref{V12}),
can be recast as:
\eq\label{I2B}
I_2=\frac{1}{2} (h_1^2+h_2^2-2h_1-4h_2-4b''b'+2L_-L_+-4a_1''a_1'-4a_2''a_2')
\,,
\en
We adopt a basis where the operators $h_j$ and $K_j$ ($j=1,2$) are diagonal,
with $K_j$ being the constituent $sp(2)$ Casimir operators defined in Eq.~(\ref{K_j_def}).

\section{Construction of UIRs in the standard mathematical literature}
\label{SSEvans}

To establish a rigorous framework for constructing the positive-energy representations,
it is essential to clarify the geometric and physical role of the Cartan subalgebra.
In our canonical $\mathfrak{so}(3,2)$ formulation under the $i$-convention,
the two mutually commuting generators $h_1$ and $h_2$ constitute the rank-2
Cartan subalgebra, serving as the fundamental operators of physical observables.
Specifically, their linear combinations dictate the maximal weights of the system,
where the operator $M_{04} = \frac{1}{2}(h_1 + h_2)$ acts as the non-perturbative
energy operator, and $L_3 = h_1 - h_2$ generates the internal spatial angular
momentum projection. 

The step-operators $A^{\pm\pm}$, connecting 
different representations of the $sp(2)\times sp(2)$ subalgebra, 
must systematically shift these Cartan weights
while strictly preserving the lowest-weight (minimal vector) constraints.
Because the standard $i$-convention scales the commutation metric down by a factor
of two compared to the historical Evans basis~\cite{Evans}, the action of the raising/lowering
root vectors shifts the eigenvalues of $h_j$ by half-integer steps. Consequently,
to ensure that the shift operators successfully map one minimal vector to another
without generating unphysical structural anomalies, the Cartan operators inside
the algebraic combinations must undergo a corresponding half-integer renormalization.
This directly necessitates the appearance of the $(h_j - \frac{1}{2})$ factors,
leading to the following precise definition of the shift operators:
\eq\label{V13} 
A^{++} &=& b'' \left(h_1-\frac{1}{2}\right) \left(h_2-\frac{1}{2}\right)
      - a_1'' L_- \left(h_2-\frac{1}{2}\right)
      - a_2'' L_+ \left(h_1-\frac{1}{2}\right)
      + a_1''a_2''b' \,, \nonumber \\
A^{+-} &=& L_+\left(h_1-\frac{1}{2}\right) - a_1''b' \,, \quad 
A^{-+} = L_-\left(h_2-\frac{1}{2}\right) - a_2''b' \,, \nonumber\\
A^{--} &=& b' \,. 
\en
In standard mathematical approaches, one considers the action of these operators exclusively
on the subspace of minimal $sp(2)\times sp(2)$ vectors, i.e., vectors $x$ satisfying
$a_j' x=0$ ($j=1,2$) which are simultaneous eigenvectors of $h_j$.
If $h_j x = \alpha_j x$, then $A^{++}x$ is a minimal eigenvector with eigenvalues 
$(\alpha_1 + \frac{1}{2}, \alpha_2 + \frac{1}{2})$; $A^{+-}x$ corresponds to 
$(\alpha_1 + \frac{1}{2}, \alpha_2 - \frac{1}{2})$; $A^{-+}x$ to 
$(\alpha_1 - \frac{1}{2}, \alpha_2 + \frac{1}{2})$; and $A^{--}x$ to 
$(\alpha_1 - \frac{1}{2}, \alpha_2 - \frac{1}{2})$. 

As in Ref.~\cite{Evans}, we require the existence of a state $e_0$ satisfying:
\eq\label{V15}
a_j'e_0=b'e_0=L_+e_0=0\,, \quad h_je_0=q_je_0 \,, \quad (j=1,2) \,,
\en
where $q_j$ are chosen to be non-zero integers. In the UIR characterized by $(q_1,q_2)$, 
the Casimir operator $I_2$ acts on all non-zero states with the eigenvalue:
\eq\label{I2C}
I_2=\frac{1}{2} (q_1^2+q_2^2-2q_1-4q_2) \,.
\en
The operators $a_1'$ and $a_2'$ lower the eigenvalues of $h_1$ and $h_2$ by two units, 
respectively, while $a_1''$ and $a_2''$ raise them by two units. Crucially, $(a_1',a_1'')$ 
leave the eigenvalues of $h_2$ invariant, and $(a_2',a_2'')$ do not alter those of $h_1$. 

For illustration, we restrict our attention to the massive, positive-energy case where 
$q_1\geq q_2>2$. Under this setup, $e_0$ represents the state of minimum energy $\mu = q_0$, 
which defines the AdS rest mass, and $s=q_1-q_2$ corresponds to the particle spin. 
We further simplify to the spinless case ($q_1=q_2=q_0$, $s=0$). The states:
\eq\label{V16} 
e_n=(A^{++})^n e_0
\en
form minimal $sp(2)\times sp(2)$ vectors with $h_1$ and $h_2$ eigenvalues equal to 
$q_0 + \frac{n}{2}$. To confirm that these states are non-vanishing for all $n=1,2,\ldots$, 
one can evaluate the norm of $e_n$. Alternatively, we can evaluate the action of 
$A^{--} A^{++}$ on $e_n$. Since this combination preserves the weight, we can write 
$A^{--} A^{++} e_n=a(n) e_n$. It follows that:
\eq\label{A-A+}
(A^{--}A^{++}-A^{++}A^{--}) e_n = [b', A^{++}] e_n = [a(n)-a(n-1)]e_n \,.
\en
Evaluating the commutator via Eqs.~(\ref{V11a})-(\ref{V11c}) yields:
\eq\label{a(n)-a(n-1)}
         [b',A^{++}]=\frac{1}{4}(h_1+h_2)\left(h_1-\frac{1}{2}\right)
         \left(h_2-\frac{1}{2}\right)
          +  (4b''b'-L_-L_+) \left(\frac{1}{2}(h_1+h_2)-1\right) 
          - \frac{1}{4}(h_1-h_2) \left(h_2-\frac{1}{2}\right) \,.
\en          
For the spinless case on the minimal subspace,
combining Eqs.~(\ref{V6}), (\ref{I2B}), 
(\ref{I2C}), and (\ref{A-A+}) leads to:
\eq\label{a(n)-a(n-1)B}
a(n)-a(n-1)=\left(n+q_0-1\right)\left[\frac{1}{2}(n+1)(n+q_0)(n+2q_0-2)
  - \frac{1}{2}n(n+q_0-2)(n-2q_0-3)\right] \,.
\en
Using the boundary condition $a(-1)=0$, integration of the difference equation yields:
\eq\label{a(n)}
a(n)=\frac{1}{2} (n+1) (n+q_0) (n+q_0-1) [n+2(q_0-1)] \,.
\en
For the massive case ($q_0>2$), $a(n)>0$ holds for all $n \geq 0$. The full basis of 
the representation space can be chosen as:
\eq\label{V19}
e(n_1,n_2,n)=(a_1'')^{n_1}(a_2'')^{n_2}e_n \quad (n,n_1,n_2=0,1,2,\ldots) \,.
\en
Since $[a_1', A^{++}]=[a_2', A^{++}]=0$, it follows from Eqs.~(\ref{scalar}) and~(\ref{a(n)}) 
that all basis vectors $e(n_1,n_2,n)$ have strictly positive norms and are mutually 
orthogonal for distinct sets of indices $(n,n_1,n_2)$. This provides a complete 
mathematical characterization of massive spinless UIRs of the AdS algebra.

\section{Physical description of spinless states in AdS theory}
\label{VS4}

In standard non-relativistic and relativistic frameworks, a spinless particle state 
is uniquely specified by three independent physical quantities (e.g., three momentum 
components, or energy, total angular momentum, and its projection). As showed above, 
the mathematical description of a spinless particle in AdS theory also depends on 
three parameters, $(n,n_1,n_2)$. While the physical interpretation of $(n_1,n_2)$ 
is clear—they correspond to the eigenvalues of the fundamental operators $h_1$ and 
$h_2$—the parameter $n$ lacks a direct physical identification since it is not 
associated with an eigenvalue of any operator in the AdS algebra. 

This issue stems from the fact that the AdS algebra has a Cartan subalgebra of rank 2, 
whereas the Galilean and Poincaré algebras possess a rank $\geq 3$. Consider two distinct 
basis states: $e(2,0,0)$ and $e(0,1,1)$. These states are mutually orthogonal and 
possess non-zero norms. However, they share the identical eigenvalues ($q_0+1$) 
for both $h_1$ and $h_2$ due to the half-integer step of the restricted representation. 

If the Conjecture presented in Sec.~\ref{Intro} were valid, there would exist a self-adjoint 
operator $B$ within the theory capable of separating these states via distinct eigenvalues. 
Yet, because the rank of the AdS algebra is strictly two, no additional independent operator 
can be constructed from the algebra generators. Consequently, the Conjecture fails in AdS 
quantum mechanics, contradicting standard relativistic intuition. This raises a foundational 
question regarding the physical relevance of the full UIR space if distinct states can 
share identical quantum numbers without any algebraic means of differentiation. 

We propose that a physical description of spinless particles in AdS theory should 
not incorporate the entire mathematical UIR space, but should be restricted to states 
uniquely distinguishable by operator eigenvalues. Specifically, we propose restricting 
the physical state space to the following subsets:
\eq
S_1 = e(0,n_1,n_2)\,, \quad S_2 = b'' e(0,n_1,n_2)\,, \quad (n_1,n_2=0,1,2,\dots) \,.
\en
Within each subset, all states are uniquely distinguished by the eigenvalues of $h_1$ and $h_2$. 
Furthermore, the states in $S_1$ and $S_2$ have opposite eigenvalue parities: if $q_0$ is even, 
the eigenvalues of $h_1, h_2$ are even in $S_1$ and odd in $S_2$ (and vice versa if $q_0$ is odd). 
Taken together, $S_1$ and $S_2$ encompass all physically realizable eigenvalues of the Cartan 
generators without introducing redundant, algebraically indistinguishable states.

\section{Phenomenological application: Thermodynamics of the modified AdS gas}
\label{Thermodynamics}

To demonstrate the physical relevance of the proposed state-space restriction,
let us investigate its consequences for the thermodynamic properties of a quantum gas
with AdS symmetry in the early Universe. The study of ideal quantum gases and field
thermodynamics in anti-de Sitter backgrounds has a long history, dating back to the
seminal work by Hawking and Page~\cite{Hawking:1982dh}, and has been extensively
developed for various bosonic systems confined within AdS boundary
geometries~\cite{Elias:2018yct}. 

In a thermal environment characterized by the inverse temperature $\beta = 1/(k_B T)$,
the thermodynamic behavior of a system is fundamentally determined by its single-particle
partition function $Z(\beta)$~\cite{Elias:2018yct}. In the framework developed in the
preceding sections under the canonical $i$-convention, the AdS energy operator is 
identified as the generator $M_{04} = \frac{1}{2}(h_1 + h_2)$, which dictates the 
spectral properties of the states.

In the standard mathematical approach reviewed in Sec.~\ref{SSEvans}
(see also Ref.~\cite{Evans}),
a particle state is defined by the full triplet of quantum numbers $(n, n_1, n_2)$.
On the basis vectors $e(n_1, n_2, n)$, the eigenvalues of the Cartan generators
$h_1$ and $h_2$ are given by $q_0 + \frac{n}{2} + 2n_1$ and $q_0 + \frac{n}{2} + 2n_2$, respectively.
Consequently, the energy spectrum takes the form:
\eq
E_{\text{std}}(n, n_1, n_2) = q_0 + \frac{n}{2} + n_1 + n_2 \,,
\en
where $n, n_1, n_2 = 0, 1, 2, \dots$. Summing over the entire unconstrained UIR space,
the standard single-particle partition function evaluates to:
\eq
Z_{\text{std}}(\beta) = \sum_{n=0}^{\infty} \sum_{n_1=0}^{\infty} \sum_{n_2=0}^{\infty}
e^{-\beta (q_0 + \frac{n}{2} + n_1 + n_2)} = \frac{e^{-q_0 \beta}}{\left(1 - e^{-\frac{\beta}{2}}\right)\left(1 - e^{-\beta}\right)^2} \,.
\label{Z_standard}
\en
In the high-temperature limit ($\beta \to 0$), Eq.~(\ref{Z_standard}) scales
as $Z_{\text{std}}(\beta) \approx 2 \beta^{-3} \sim T^3$.
This behavior is consistent with standard conformal scaling laws for fields
in a bulk $(3+1)$-dimensional space, where the partition function effectively
samples three continuous spatial degrees of freedom~\cite{Hawking:1982dh,Elias:2018yct}.

In contrast, our physical framework restricts the admissible states to the subsets
$S_1$ and $S_2$. For the subset $S_1 = e(0, n_1, n_2)$, the unphysical quantum number
is frozen at $n=0$, yielding the energy spectrum:
\eq
E_1(n_1, n_2) = q_0 + n_1 + n_2 \,.
\en
The partition function for the $S_1$ sector is then given by:
\eq
Z_{S_1}(\beta) = \sum_{n_1=0}^{\infty} \sum_{n_2=0}^{\infty} e^{-\beta (q_0 + n_1 + n_2)}
= \frac{e^{-q_0 \beta}}{(1 - e^{-\beta})^2} \,.
\en
For the second subset, $S_2 = b'' e(0, n_1, n_2)$, the parity-changing operator
$b''$ satisfies $[h_j, b''] = b''$, thereby shifting the eigenvalues of both
$h_1$ and $h_2$ upward by one unit. Under the $i$-convention, this shifts the total 
energy eigenvalue $M_{04} = \frac{1}{2}(h_1+h_2)$ by exactly one unit ($\Delta E = 1$), 
resulting in the spectrum:
\eq
E_2(n_1, n_2) = q_0 + n_1 + n_2 + 1 \,.
\en
The partition function for the $S_2$ sector reads:
\eq
Z_{S_2}(\beta) = \sum_{n_1=0}^{\infty} \sum_{n_2=0}^{\infty} e^{-\beta (q_0 + n_1 + n_2 + 1)}
= \frac{e^{-(q_0 + 1) \beta}}{(1 - e^{-\beta})^2} \,.
\en
Combining the two physically realizable sectors, the total modified partition function is given by:
\eq
Z_{\text{mod}}(\beta) = Z_{S_1}(\beta) + Z_{S_2}(\beta) = e^{-q_0 \beta}
\frac{1 + e^{-\beta}}{(1 - e^{-\beta})^2} \,.
\label{Z_modified}
\en
To evaluate the macroscopic impact of this modification,
we calculate the Helmholtz free energy of the single-particle system,
defined as $F = -k_B T \ln Z = -\frac{1}{\beta} \ln Z$.
For the standard and modified cases, we find:
\eq\label{F_std}
F_{\text{std}} = q_0 + \frac{1}{\beta}
\ln\left(1 - e^{-\frac{\beta}{2}}\right) + \frac{2}{\beta} \ln\left(1 - e^{-\beta}\right) \,,
\en
\eq\label{F_mod}
F_{\text{mod}} = q_0 - \frac{1}{\beta}
\ln\left(1 + e^{-\beta}\right) + \frac{2}{\beta}
\ln\left(1 - e^{-\beta}\right) \,.
\en
In the high-temperature regime ($\beta \to 0$), the free energies behave asymptotically as:
\eq
F_{\text{std}} \approx -\frac{3}{\beta} \ln \left(\frac{1}{\beta}\right) \sim -3 k_B T \ln(T) \,,
\en
\eq
F_{\text{mod}} \approx -\frac{2}{\beta} \ln \left(\frac{1}{\beta}\right) \sim -2 k_B T \ln(T) \,.
\en
To fully understand the macroscopical implications of this
algebraic state-space restriction, it is instructive to evaluate
both the high-temperature ($T \to \infty$, $\beta \to 0$) and
low-temperature ($T \to 0$, $\beta \to \infty$) asymptotic limits
for the thermodynamic quantities under the canonical $i$-convention.

In the high-temperature limit ($\beta \to 0$), the exponential factors can be
approximated via the standard Taylor expansion, $e^{-x} \approx 1 - x$.
Applying this to the Helmholtz free energies derived in Eqs.~(34) and (35),
we obtain the precise asymptotic expansions:
\eq
F_{\text{std}}(\beta \to 0) &\approx& \frac{1}{\beta} \ln \left(\frac{\beta}{2}\right)
+ \frac{2}{\beta} \ln(\beta) \approx -\frac{3}{\beta} \ln \left(\frac{1}{\beta}\right)
\sim - 3 k_B T \ln(T) \,, \label{F_std_high} \\
F_{\text{mod}}(\beta \to 0) &\approx& -\frac{1}{\beta} \ln(2)
+ \frac{2}{\beta} \ln(\beta) \approx -\frac{2}{\beta}
\ln \left(\frac{1}{\beta}\right) \sim -2 k_B T \ln(T) \,. \label{F_mod_high}
\en
Differentiating the free energy with respect to the inverse temperature
yields the internal energy of the systems,
$U = \frac{\partial(\beta F)}{\partial \beta}$.
From Eqs.~(\ref{F_std_high}) and (\ref{F_mod_high}),
the internal energy asymptotes at high temperatures read:
\eq
U_{\text{std}} \approx 3 k_B T \,, \quad U_{\text{mod}} \approx 2 k_B T \,.
\en
This comparison explicitly demonstrates the thermodynamic impact of
the rank-2 Cartan algebra restriction. The full mathematical UIR
samples three continuous spatial degrees of freedom, yielding the
standard $U \sim 3k_B T$ scaling inherent to a generic $(3+1)$-dimensional
conformal bulk field theory~\cite{Hawking:1982dh,Elias:2018yct}.
In sharp contrast, freezing the redundant topological parameter at $n=0$
effectively removes one topological degree of freedom from the physical spectrum,
resulting in a thermodynamically stiffer system governed by a reduced
$U \sim 2k_B T$ equation of state.

Conversely, in the deep low-temperature limit ($\beta \to \infty$),
the thermal fluctuations freeze out, and the partition functions are
dominated exclusively by the lowest-lying vacuum energy configurations.
Expanding Eqs.~(28) and (33) for large $\beta$ fields yields:
\eq
Z_{\text{std}}(\beta \to \infty) &\approx& e^{-q_0 \beta} \left(1 + e^{-\frac{\beta}{2}} + 2e^{-\beta} + \dots \right) \,, \\
Z_{\text{mod}}(\beta \to \infty) &\approx& e^{-q_0 \beta} \left(1 + e^{-\beta} + 2e^{-2\beta} + \dots \right) \,.
\en
The corresponding low-temperature Helmholtz free energies collapse to:
\eq
F_{\text{std}}(\beta \to \infty) \approx q_0 - \frac{1}{\beta} e^{-\frac{\beta}{2}} \,, \quad 
F_{\text{mod}}(\beta \to \infty) \approx q_0 - \frac{1}{\beta} e^{-\beta} \,.
\en
This highlights a crucial physical signature: at zero temperature
($T \to 0$, $\beta \to \infty$), both theories smoothly converge
to the identical exact vacuum rest energy $F = q_0$, as dictated
by the lowest-weight state $e_0$. However, the path toward the
vacuum state differs fundamentally. The standard UIR partition
function receives unphysical soft thermal corrections from the
first topological excitation step $\Delta E = 1/2$ associated
with the $n=1$ mode. In our restricted framework, because
all $n > 0$ configurations are eliminated to preserve strict
operator observability, the first allowed thermal excitation
in $S_2$ requires a full energy gap of $\Delta E = 1$.
This implies that the restricted AdS quantum gas remains
protected against low-temperature unphysical state mixtures,
revealing a much tighter phase-space security bound than
standard geometric field-theoretic setups.

This result exhibits a highly distinct cosmological signature.
Eliminating the redundant quantum number $n$ effectively freezes
out one intrinsic topological degree of freedom inherent to the
full AdS UIR. Consequently, a quantum gas governed
by this modified symmetry becomes thermodynamically stiffer,
yielding a lower heat capacity and an altered equation of state
($U_{\text{mod}} \approx 2 k_B T$ vs. $U_{\text{std}} \approx 3 k_B T$).
In the context of early Universe cosmology, such a modification
directly accelerates the cooling rate during the radiation-dominated
era, potentially leaving observable imprints on the primordial gravitational
wave spectrum and cosmic microwave background (CMB) anisotropies.

\section{Conclusion}
\label{Conclusion}

In this paper, we demonstrated that the quantum description of elementary particles under
AdS symmetry departs significantly from standard non-relativistic and relativistic paradigms.
Because the Cartan subalgebra of the AdS algebra is of rank 2, a physically consistent
formulation requires restricting the state space to specific sets of vectors uniquely
characterized by operator eigenvalues ($S_1$ and $S_2$), rather than utilizing
the full mathematical UIR space. We showed that the standard quantum-mechanical
assumption—that distinct states can always be distinguished by a self-adjoint operator—fails
within an exact AdS framework due to the presence of the unphysical parameter $n$.

The thermodynamic analysis presented in Sec.~\ref{Thermodynamics} shows that this state-space
restriction has immediate, dramatic consequences for the physics of the early Universe,
altering the single-particle partition function from a $T^3$ dependence to a stiffer $T^2$
behavior at high temperatures. 

In future work, we plan to extend these algebraic insights
to other major phenomenological areas:

\begin{itemize}

\item {\bf Modification of Hadronic Form Factors in AdS/QCD:}
  
  In holographic QCD models, such as
  the hard-wall~\cite{Ehrlich:2005qh} or soft-wall~\cite{Karch:2006pv}
  approaches, hadronic wave functions are conventionally expanded over
  a complete set of bulk modes to evaluate electromagnetic and transition
  form factors~\cite{Brodsky:2014cda}-\cite{Lyubovitskij:2023lrp}.
  Restricting the admissible Hilbert space to $S_1 \cup S_2$ will modify
  the underlying completeness relations and sum rules.
  This is expected to directly alter the functional behavior
  of the electromagnetic form factors of mesons and nucleons
  at intermediate momentum transfers $Q^2$,
  providing a clean test of the algebraic restriction.

\item {\bf Regge Trajectories and Hadronic Spectroscopy:}
 
  In standard semiclassical and holographic
  approaches, radial excitations are typically mapped
  onto geometric coordinates or sharp infrared
  cutoffs to reproduce the linear behavior of hadron
  masses~\cite{Ehrlich:2005qh}-\cite{Lyubovitskij:2023lrp}.
  By treating the problem purely algebraically and removing
  the redundant parameter $n$, the mass spectra
  of light mesons $M^2$ will depend uniquely
  on the $h_1$ and $h_2$ operators.
  This invariant setup can shed light on the splitting
  and slopes of Regge trajectories for
  radially excited states without introducing
  ad-hoc geometric boundaries or artificial potentials.

\end{itemize}

In summary, transitioning from standard QFT representations to a physically motivated,
restricted algebraic framework provides a new avenue for resolving foundational issues
while offering distinct signatures for particle phenomenology and early Universe cosmology.

\begin{acknowledgments} 

  This work was funded by FONDECYT (Chile)
  under Grants No.~1230160 and No.~1240066, and by the ANID$-$Millennium
  Science Initiative Program$-$ICN2019\_044 (Chile).

\end{acknowledgments}

\end{document}